\documentclass[a4paper,onecolumn,draft]{IEEEtran}
\usepackage{amssymb,stmaryrd,amsmath,amsfonts,rotating,mathrsfs,amsfonts}
\usepackage[noadjust]{cite}
\usepackage{color}
\usepackage{epic}
\RequirePackage{bbm}
\newtheorem{theorem}{Theorem}

\newcommand{\btheo}{\begin{theorem}}
\newcommand{\etheo}{\end{theorem}}
\newcommand{\bproof}{\begin{proof}}
\newcommand{\eproof}{\end{proof}}
\newtheorem{definition}[theorem]{Definition}
\newcommand{\bdefi}{\begin{definition}}
\newcommand{\edefi}{\end{definition}}
\newtheorem{fact}[theorem]{Fact}
\newcommand{\bprop}{\begin{fact}}
\newcommand{\eprop}{\end{fact}}
\newtheorem{corollary}[theorem]{Corollary}
\newcommand{\bcor}{\begin{corollary}}
\newcommand{\ecor}{\end{corollary}}
\newtheorem{example}[theorem]{Example}
\newcommand{\bex}{\begin{example}}
\newcommand{\eex}{\end{example}}
\newtheorem{lemma}[theorem]{Lemma}
\newcommand{\blemma}{\begin{lemma}}
\newcommand{\elemma}{\end{lemma}}
\newtheorem{remark}[theorem]{Remark}
\newcommand{\bremark}{\begin{remark}}
\newcommand{\eremark}{\end{remark}}
\newtheorem{conj}[theorem]{Conjecture}
\newcommand{\bconj}{\begin{conj}}
\newcommand{\econj}{\end{conj}}

\newcommand{\naturals}{\ensuremath{\mathbb{N}}}

\def\0{{\tt 0}} 
\def\1{{\tt 1}} 
\def\?{{\tt *}} 
\renewcommand{\mid}{\,|\,}

\renewcommand{\dh}{{{d_{H}}}}

\allowdisplaybreaks
\begin{document}
\title{Rate-Dependent Analysis of the Asymptotic Behavior of Channel Polarization }
%
\author{S.~Hamed~Hassani,~\IEEEmembership{Student~Member,~IEEE,} Ryuhei~Mori,~\IEEEmembership{Student~Member,~IEEE,} Toshiyuki~Tanaka,~\IEEEmembership{Member,~IEEE,} and~R\"{u}diger~Urbanke%
\thanks{ The material in
this paper was presented in part in \cite{M10}, \cite{MT102}, \cite{HU10} and \cite{T10}.}%
\thanks{The work of S. H. Hassani was supported by grant no 200021-121903 of the Swiss National Science Foundation,
the work of R. Mori by the Grant-in-Aid for Scientific Research for JSPS Fellows (22$\cdot$5936), MEXT, Japan,
and the work of T. Tanaka by the Grant-in-Aid for Scientific Research (C) (22560375), JSPS, Japan.
}
\thanks{S. H. Hassani and R. Urbanke are with the School of Computer and Communication Science,
EPFL, CH-1015 Lausanne, Switzerland
(e-mail: \{seyehamed.hassani, rudiger.urbanke\}@epfl.ch).}%
\thanks{R. Mori and T. Tanaka are with the Department of Systems Science, Graduate School of Informatics, 
Kyoto University, Yoshida Hon-machi, Sakyo-ku, Kyoto-shi, Kyoto, 606-8501 Japan
(e-mail: rmori@sys.i.kyoto-u.ac.jp, tt@i.kyoto-u.ac.jp).}%
}

\maketitle

\begin{abstract}
For a binary-input memoryless symmetric channel $W$, we consider the asymptotic behavior of the polarization process in the large block-length regime
when transmission takes place over $W$. In particular, we study the
asymptotics  of the  cumulative distribution $\mathbb{P}(Z_n \leq z)$,
where $\{Z_n\}$ is the Bhattacharyya process defined from $W$, and its dependence on
the rate of transmission. On the basis of this result, we characterize the asymptotic
behavior, as well as its dependence on the rate, of the block error probability of polar codes using the
successive cancellation decoder. This refines the original bounds
by Ar{\i}kan and Telatar. Our results apply to general polar codes based
on $\ell \times \ell$  kernel matrices.

We also provide lower bounds on the block error
probability of polar codes using the MAP decoder. The MAP lower bound
and the successive cancellation upper bound coincide when $\ell=2$,
but there is a gap for $\ell>2$.  \end{abstract}

\section{Introduction}
\subsection{Polar Codes}
\IEEEPARstart{P}{olar} codes, introduced by Ar\i kan \cite{Ari09}, are a family of codes
that provably achieve the capacity of binary-input memoryless symmetric (BMS)
channels using low-complexity encoding and decoding algorithms. Since their invention, there has been a large body of work that has analyzed (see e.g., \cite{AT09}--\cite{T10})  and extended 
 (see e.g., \cite{KU} --\cite{AT}) 
these codes.

The construction of polar codes is based on an $\ell \times \ell$ matrix
$G$, with entries in $\{0,1\}$, called the kernel matrix. Besides being invertible, the matrix $G$ should have the
property that  none of its column permutations is upper triangular~\cite{KSU}.  We call a matrix $G$ 
with such properties a \emph{polarizing} matrix and in the following, whenever we speak of a kernel matrix $G$, we assume that  $G$ is polarizing. 

The rows of the generator matrix of a polar code with block-length $N= \ell^n$
are chosen from the rows of the matrix 
\begin{equation*}
G^{\otimes n}\triangleq\overbrace{G\otimes G\otimes\cdots\otimes G}^{n},
\end{equation*}
where $\otimes$ denotes the Kronecker product. For the case $\ell=2$ and the
choice $G=\bigl[ \begin{smallmatrix} 1 &0 \\ 1& 1 \end{smallmatrix}
\bigr]$, Reed-Muller (RM) codes also fall into this category.  However,
the crucial difference between polar codes and RM codes lies in the
choice of the rows. For RM codes, the rows of the largest
weights are chosen, whereas for polar codes the choice is dependent on the
channel and is made using a method called channel polarization. We briefly review this method
 and explain how polar codes are constructed from it. We also refer the reader to \cite{Ari09}, \cite{Kor09thesis}  and  \cite{KSU} for
a detailed discussion. 

\subsection{Channel Polarization}
Let $W$ be a BMS channel, and let $\mathcal{X}=\{0,1\}$ denote its input alphabet, 
$\mathcal{Y}$ the output alphabet, and  $W(y \mid x)$ the transition probabilities. 
Let $I(W) \in [0,1]$ denote the mutual information between the input and output of $W$ with uniform distribution on the input. The capacity of a BMS channel $W$ is equal to $I(W)$. 
Also,  the Bhattacharyya parameter of $W$, denoted by $Z(W)$,
is defined as
\begin{align*}
& Z(W)= \sum_{y \in \mathcal{Y}} \sqrt{W(y\mid 0)W(y \mid 1)}.
\end{align*}
It provides upper and lower bounds of the error probability $P_e(W)$
in estimating the channel input $x$ on the basis of 
the channel output $y$ via the maximum-likelihood (ML)
decoding of $W(y|x)$ as follows~\cite[Chapter 4]{RiU08}, \cite{Kor09thesis}.
\begin{equation}\label{eq:PeZ}
\frac12\left(1-\sqrt{1-Z(W)^2}\right)\le P_e(W)\le \frac12 Z(W).
\end{equation}
It is also related to the capacity $I(W)$ via
\begin{align*}
&Z(W)+I(W)\ge1,\\
&[Z(W)]^2+[I(W)]^2\le1,
\end{align*}
both proved in~\cite{Ari09}. 

The method of channel polarization is defined as follows. 
Take $N=\ell^n$ copies of a BMS channel $W$.
Combine them by using the kernel matrix $G$ to make a new set of $ \ell^n$
channels $\{W_{ \ell^n} ^{(i)}\}_{1 \leq i \leq  \ell^n}$.  The construction of
these channels is done by recursively applying a transform called channel
splitting. Channel splitting is a transform which takes a BMS channel $W$
as input and outputs $\ell$ BMS channels $W ^j $, $0 \leq j \leq \ell-1$. 
The channels $W ^j $ are constructed according to the following rule:   
Consider a random row vector $U_0^{\ell-1}=(U_0,\,\ldots,\,U_{\ell-1})$ that is uniformly distributed over $\{0, 1\}^{\ell}$. 
Let $X_0^{\ell-1}=U_0^{\ell-1} G$, where the arithmetic is in $\mathrm{GF}(2)$. 
Also, let $Y_0^{\ell-1}$ be the output of $\ell$ uses of $W$ over the input $X_0^{\ell-1}$. We define the
 channel between $U_0^{\ell-1}$ and $Y_0^{\ell-1}$ by the transition probabilities
\begin{equation}
W_\ell(y_0^{\ell-1} \mid u_0^{\ell-1}) \triangleq \prod_{i=0}^{\ell-1} W(y_i \mid x_i)=\prod_{i=0}^{\ell-1} W(y_i\mid (u_0^{\ell-1} G)_i).
\end{equation}
 The channel $W^j :  \{0, 1\} \to \mathcal{Y}^{\ell} \times \{0, 1\}^{j}$ is defined as the BMS channel 
with input $u_j$, output $(y_0^{\ell-1},u_0^{j-1})$ and transition probabilities
\begin{equation}
W^j(y_0^{\ell-1}, u_0^{j-1} \mid u_j)= \frac{1}{2^{\ell-1}}\sum_{u_{j+1}^{\ell-1}} W_\ell(y_0^{\ell-1} \mid u_0^{\ell-1}).
\end{equation}
Here and hereafter, $u_i^j$ denotes the subvector $(u_i,\,\ldots,\,u_j)$. 

The construction of the channels $\{W_{ \ell^n} ^{(i)}\}_{1 \leq i \leq  \ell^n}$ can be visualized
in the following way~\cite{Ari09}. Consider an infinite $\ell$-ary
tree with the root node placed at the top. 
To each vertex of the tree, we assign a channel in a way that the
collection of all the channels that correspond to the vertices at depth
$n$ equals $\{W_{ \ell^n} ^{(i)}\}_{1 \leq i \leq  \ell^n}$.  We do
this by a recursive procedure. Assign to the root node the channel $W$
itself. From left to right, assign $W^0$ to $W^{\ell-1}$ to the children
of the root node.  In general, if $Q$ is the channel that is assigned
to vertex $v$, we assign $Q^0$ to $Q^{\ell-1}$, from left to right respectively, 
to the children of the node $v$. 
There are $ \ell^n$ vertices at level $n$ in this $\ell$-ary tree.
Assume that we label these vertices from left to right from $1$ to $
\ell^n$. Let the channel assigned to the $i$th vertex, $1
\leq i \leq \ell^n$, be $W_{ \ell^n} ^{(i)}$.  Also, let the $\ell$-ary representation
of $i-1$ be $b_1 b_2 \cdots b_n$, where $b_1$ is the most significant
digit. Then we have
\begin{equation*}
 W_{ \ell^n} ^{(i)}=(((W^{b_1})^{b_2})^{\cdots})^{b_n}.
\end{equation*}
As an example, assuming $i=7$, $n=3$ and $\ell=2$ we have $W_{8} ^{(7)} =((W^{1})^{1})^{0}$. 

The channels $\{W_{ \ell^n} ^{(i)}\}_{1 \leq i \leq  \ell^n}$ have the property that,
 as $n$ grows large, a fraction close to $I(W)$
of the channels have capacity close to $1$ (or Bhattacharyya parameter close to $0$); and a fraction close to $1-
I(W)$ of the channels have capacity close to $0$ (or Bhattacharyya parameter close to $1$).  
The basic idea behind polar codes is to use those channels with capacity close to $1$ 
for information transmission. 
Accordingly, given the rate $R<I(W)$ and block-length $N=\ell^n$, the rows 
of the generator matrix of a polar code of block-length $N$ correspond to a subset of the rows of the matrix $G^{\otimes n}$ whose indices 
are chosen with the following rule:  Choose a subset of size $NR$ of the channels
$\{W_{ \ell^n} ^{(i)}\}_{1 \leq i \leq  \ell^n}$  with  the least values for the Bhattacharyya parameter and choose 
the rows $G^{\otimes n}$ with the indices corresponding to those of the channels. 
For example, if the channel $W_{\ell^n}^{(i)}$ is chosen, 
then the $j$th row of $G^{\otimes n}$ is selected, 
where the $\ell$-ary representation of $j-1$ is the digit-reversed version of that of $i-1$. 
We decode using a successive cancellation (SC) decoder. This algorithm decodes the bits one-by-one in 
a pre-chosen order that is closely related to how the row indices of $G^{\otimes n}$ are chosen.

\subsection{Problem Formulation and Relevant Work} \label{related}
Let $\mathcal{I}$ be the set of indices of the $NR$ channels in the set
$\{W_{ \ell^n} ^{(i)}\}_{1 \leq i \leq  \ell^n}$  with the least values for the Bhattacharyya parameter. 
Let $\mathbb{P}_e^{\text{SC}}(N,R)$ and $\mathbb{P}_e^{\text{MAP}}(N,R)$ denote 
the average block
error probability of the SC and the maximum a-posteriori (MAP) decoders, respectively, 
with block-length $N$ and rate $R$.
For the SC decoder we have~\cite{Ari09, Kor09thesis},
\begin{equation}\label{eq:ineqZ}
 \max_{i \in \mathcal{I}} \frac 12 \left(1- \sqrt{1-Z(W_{ \ell^n} ^{(i)})^2}\right)  \leq \mathbb{P}_e^{\text{SC}}(N,R)
\leq \sum_{i \in \mathcal{I}} Z(W_{ \ell^n} ^{(i)}).
\end{equation}
This relation evidently shows that the distribution of the Bhattacharyya parameters of the channels $\{W_
{ \ell^n} ^{(i)}\}_{1 \leq i \leq  \ell^n}$ plays a fundamental role
in the analysis of polar codes. More precisely, for $n \in \naturals\triangleq\{0,\,1,\,2,\,\ldots\}$
and $0 < z < 1$, we are interested in analyzing the behavior of 
\begin{equation}\label{equ:1}
F(n, z) = \frac{\# \{ i : Z(W_{ \ell^n} ^{(i)}) \leq z \}}{ \ell^n},
\end{equation}
where $\#A$ denotes the number of elements of the set $A$. 
There is an entirely equivalent probabilistic description of
\eqref{equ:1}: Define the ``polarization''  process~\cite{AT09}
of the channel $W$ as a channel-valued stochastic process $\{W_n\}_{n\in\naturals}$ with $W_0=W$ and
\begin{equation}
\label{eq:pol-proc}
W_{n+1}=  W_n^{B_n},
\end{equation}
where $\{B_n\}_{n \in \naturals}$ is a sequence of independent and 
identically-distributed (i.i.d.) random variables
with distribution $\mathbb{P}(B_0=j)=\frac{1}{\ell}$ for $j\in\{0,\,1,\,\ldots,\,
\ell-1\}$. In other words, the process begins at the root node of the
infinite $\ell$-ary tree introduced above, and in each step it chooses one of the $\ell$
children of the current node with uniform probability.  So at time $n$,
the process $\{W_n\}_{n\in\naturals}$ outputs one of the $ \ell^n$ channels at level $n$ of
the tree uniformly at random. The Bhattacharyya process $\{Z_n\}_{n\in\naturals}$ of the channel
$W$ is defined from the polarization process as $Z_n \triangleq Z(W_n)$. 
In this setting we have
\begin{equation}
 \mathbb{P}(Z_n \leq z) = F(n,z).
\end{equation}
It was shown in \cite{AT09} and \cite{Kor09thesis} that the Bhattacharyya 
process $\{Z_n\}_{n\in\naturals}$ 
 converges almost surely to a $\{0,1\}$-valued random variable
$Z_{\infty}$ with $\mathbb{P}(Z_{\infty}=0)=I(W)$. Our objective is to
investigate the asymptotic behavior of  $\mathbb{P}(Z_n \leq z)$. The analysis of
the process $\{Z_n\}_{n\in\naturals}$ around the point $z=0$ is of particular interest, as
this indicates how the ``good'' channels (i.e., the channels that have
mutual information close to $1$) behave.  The asymptotic analysis of the
process is closely related to the ``partial distances'' of the kernel
matrix $G$: 
\begin{definition}[Partial Distances]\label{def:partial-d}
We define the \emph{partial distances} $D_i(G)$, $i=0,\cdots, \ell-1$, of an $\ell
\times \ell$ matrix
$G= \biggl[ \begin{smallmatrix} g_{0} \\  \vdots \\ g_{\ell-1} \end{smallmatrix}
\biggr]$ ($g_i$'s are row vectors)
as
\begin{align*}
D_i(G) & \triangleq \dh(\{g_i\},\langle g_{i+1},\dotsc,g_{\ell-1}\rangle), \qquad i=0,\dotsc,\ell-2, \\
D_{\ell-1}(G) & \triangleq \dh(\{g_{\ell-1}\},\{0\}),
\end{align*}
where $d_H(\cdot, \cdot)$ denotes the Hamming distance between two sets of binary sequences, 
and where $\langle g_{i+1},\dotsc,g_{\ell-1}\rangle$ denotes the linear space 
spanned by $g_{i+1},\,\dotsc,\,g_{\ell-1}$.
The \textit{exponent} of $G$  is then defined as 
\begin{align*}
E(G) = \frac{1}{\ell}\sum_{i=0}^{\ell-1} \log_\ell D_i(G),
\end{align*}
and the \textit{second exponent} of $G$ is defined as
\begin{align*}
V(G) = \frac{1}{\ell}\sum_{i=0}^{\ell-1} (\log_\ell D_i(G) - E(G))^2.
\end{align*}
\end{definition}
In other words, the exponent $E(G)$ and the second exponent $V(G)$ 
are the mean and the variance of the random variable $\log_\ell D_B(G)$, 
where $B$ is a random variable taking a value in $\{0,\,1,\,\ldots,\,\ell-1\}$ 
with uniform probability.
It should be noted that the invertibility of $G$ implies the partial distances $\{D_i(G)\}$ 
to be strictly positive, making the exponent $E(G)$ finite.
Note also that the condition for a matrix $G$ to be polarizing, 
that none of column permutations of $G$ is upper triangular, 
implies $\{D_i(G)\}$ to be strictly greater than 1, yielding $E(G)$ to be 
strictly positive.

The following theorem partially characterizes the behavior of the process $\{Z_n\}_{n\in\naturals}$ around $z=0$.
\begin{theorem}[\cite{AT09} and \cite{Kor09thesis}] \label{AT}
Let $W$ be a BMS channel and assume that we are using as the kernel matrix an $\ell\times\ell$ matrix $G$ with exponent $E(G)$. For any fixed $\beta$ with $0 < \beta < E(G)$,
\begin{equation*}
 \lim_{n \to \infty}\mathbb{P}(Z_n \leq 2^{-\ell^{n \beta}}) = I(W).
\end{equation*}
Conversely, if $I(W)<1$, then for any fixed $\beta > E(G)$,
\begin{equation*}
 \lim_{n \to \infty}\mathbb{P}(Z_n \geq 2^{-\ell^{n \beta}}) = 1.
\end{equation*}
\IEEEQED
\end{theorem}
An important consequence of Theorem~\ref{AT} is that, as the behavior of
$\mathbb{P}_e ^{\text{SC}}(N,\,R)$ when using polar codes
with the kernel matrix $G$, of block-length $N= \ell^n$ and rate
$R<I(W)$ under SC decoding is asymptotically the same as that of $\max_{i\in\mathcal{I}}Z(W_N^{(i)})$ from~\eqref{eq:ineqZ},
the probability of error behaves as
$2^{-\ell^{nE(G) + o(n)}}$ as $N$ tends to infinity. 
A noteworthy point about this result is that the asymptotic analysis 
of the probability of error is rate-independent, provided that 
the rate $R$ is less than the capacity $I(W)$. 
In this paper,
we provide a refined estimate for $\mathbb{P}(Z_n \leq z)$. Specifically, we derive
the asymptotic relation between  $\mathbb{P}(Z_n \leq z)$ and the rate
of transmission $R$. From this we derive the asymptotic behavior of
$\mathbb{P}_e ^{\text{SC}}(N,\,R)$ and its dependence on the rate of transmission. We further derive
lower bounds on the error probability when we perform MAP decoding instead of SC decoding.  

An important point to mention here is that the results of this paper are obtained in the asymptotic
 limit of the block-length for any \emph{fixed} rate value $R$. Considering the 
regime where $R$ also varies with the block-length is a problem of different interest, 
for which we refer the reader to~\cite{AHKU}.   

The outline of the paper
is as follows. In Section~\ref{s_r} we state the main results of the
paper. In Section~\ref{a_p} we first define several auxiliary processes
and provide bounds on their asymptotic behavior. Using these bounds,
we then prove the main results. We discuss the implications of the proofs in selecting the set of channel indices in 
Section~\ref{select_sec}. 
It should be noted that in the following the logarithms
are in base 2 unless explicitly stated otherwise.

\section{Main Results}\label{s_r}
\begin{theorem}\label{main_result}
Consider an $\ell \times \ell$ polarizing kernel matrix $G= \biggl[ \begin{smallmatrix} g_{0} \\  \vdots \\ g_{\ell-1} \end{smallmatrix}
\biggr]$. For a BMS channel $W$, let $\{Z_n= Z(W_n)\}_{n\in\naturals}$ be the Bhattacharyya process of $W$.
Let $Q(t)\triangleq\int_t^\infty e^{-z^2/2}\,dz/\sqrt{2\pi}$ be the error function 
and $Q^{-1}(\cdot)$ be its inverse function. 
\begin{enumerate}
\item For $R < I(W)$,
\begin{equation} \nonumber
 \lim _{n \to \infty} \mathbb{P}\left(Z_{n} \leq 2^{ -\ell^{nE(G)+ \sqrt{nV(G)} Q^{-1} \left(\frac{R}{I(W)}\right) + f(n) }}\right) =R.
\end{equation}
\item Let
$H = [g_{\ell-1}^{\rm T},\,\cdots,\,g_{0}^{\rm T}]^{-1}$ ($\cdot^{\rm T}$ denotes the transpose) and assume that $D_{i}(H) \leq {D_{i-1}(H) }$ for $1 \leq i \leq \ell-1$. Then, for $R' < 1- I(W)$ we have,
\begin{equation} \nonumber
 \lim _{n \to \infty} \mathbb{P}\left(Z_{n} \geq 1- 2^{ -\ell^{nE(H)+ \sqrt{nV(H)} Q^{-1} \left(\frac{R'}{1-I(W)}\right) + f(n) }}\right) =R'.
\end{equation}
\end{enumerate}
Here, $f(n)$ is any function satisfying $f(n) = o(\sqrt{n})$. 
\IEEEQED
\end{theorem}
{\em Discussion:} Theorem~\ref{main_result} characterizes the asymptotic behavior of $\mathbb{P}(Z_n \leq z)$ and
refines Theorem~\ref{AT} in the following way. According to Theorem~\ref{AT}, if we transmit at rate $R$ below the channel capacity, 
then the quantity $ \log_\ell(-\log(\mathbb{P}_{e} ^{\text{SC}}(N=\ell^n,\,R)))$ scales like $nE(G) + o(n)$. The first part of Theorem~\ref{main_result} gives one further 
term by stating that $o(n)$ is in fact $\sqrt{nV(G)} Q^{-1}\left(\frac{R}{I(W)}\right)+ o(\sqrt{n})$.  
The second part of Theorem~\ref{main_result}, on the other hand, characterizes the asymptotic behavior 
of $\mathbb{P}(Z_n\leq z)$ near $z=1$, which is important in applications of polar codes for source coding \cite{KU}. 
Put together, Theorem~\ref{main_result} characterizes the scaling of the error probability of polar codes with the SC decoder. Similar results hold 
for the case of the MAP decoder.
\begin{theorem} \label{main_result2}
Let $W$ be a BMS channel and let $R < I(W)$ be the rate of transmission. Consider an $\ell \times \ell$ kernel matrix $G$ with $\{w_0(G), \cdots, w_{\ell-1}(G) \}$ the Hamming weights 
of its rows and 
define
\begin{equation}
 E_w(G)=\frac{1}{\ell} \sum_{i=0}^{\ell-1} \log_\ell w_i(G), \quad V_w(G)= \frac{1}{\ell}\sum_{i=0}^{\ell-1} (\log_\ell w_i(G)-E_w(G))^2.
\end{equation}
If we use polar codes of length $N=\ell^n$ and rate $R$ for transmission, then 
the probability of error under MAP decoding,  $\mathbb{P}_e^{\text{MAP}}(N,R)$,  satisfies 
\begin{equation}
 \log_\ell (-\log (\mathbb{P}_e^{\text{MAP}}(N,R))) \leq n E_w(G) + \sqrt{n V_w(G)}Q^{-1} \left(\frac{R}{I(W)}\right) + o(\sqrt{n}).
\end{equation}
\IEEEQED
\end{theorem}
{\em Discussion:} Let $G$ be according to Ar{\i}kan's original construction~\cite{Ari09}, 
i.e., $G= \bigl[ \begin{smallmatrix} 1 & 0 \\ 1 & 1  \end{smallmatrix} \bigr]$, 
which is the only polarizing matrix for the case $\ell=2$.
For this $G$, we have $w_i(G) = D_i(G)$ for $i = 0$ and $1$. Hence, the block
error probability for the SC decoder and the MAP block error
probability share the same asymptotic behavior according to Theorems~\ref{main_result} and \ref{main_result2}. 
For a general $\ell \times \ell$ matrix $G$, however, one
may have strict inequality $E_w(G) > E(G)$, in which case one
still has an asymptotic  gap between the  
error probability with SC decoding and the lower bound of
MAP error probability. Whether or not this gap can be filled 
or made narrower is an open problem.

\section{Proof of the Main Result} \label{proofs} \label{a_p}

\subsection{Preliminaries}

Let $\{B_n\}_{n \in \naturals}$ be a sequence of i.i.d. random variables that take their 
values in $\{0,1, \cdots, \ell-1\}$ with uniform probability, i.e.,
 $\mathbb{P}(B_0=j) = \frac{1}{\ell}$ for $j \in \{0,\,1,\,\ldots,\, \ell -1\}$. 
Let $( \Omega, \mathcal{F}, \mathbb{P})$ denote the probability space generated by 
the sequence $\{B_n\}_{n\in\naturals}$ and let  $(\Omega_n, \mathcal{F}_n, \mathbb{P}_n)$ be the probability space generated 
by $(B_0, \cdots,B_n)$. We now couple the polarization process $\{W_n\}_{n\in\naturals}$ 
with the sequence $\{B_n\}_{n \in \naturals}$ via~\eqref{eq:pol-proc}.
Consequently, the Bhattacharyya process $\{Z_n = Z(W_n)\}_{n\in\naturals}$ is coupled with the sequence $\{B_n\}_{n\in\naturals}$.
By using the bounds given in \cite[Chapter 5]{Kor09thesis} we have the
following relationship between the Bhattacharyya  parameters of  $W
^i$ and that of $W$: Recall that $\{D_i(G)\}_{0 \leq i \leq \ell-1}$ are the partial distances of the matrix $G$. 
We have~\cite{Kor09thesis}
\begin{align} \label{general_Z(G)}
& Z(W)^{D_{i}(G)} \le Z(W^{i}) \le 2^{\ell-i}Z(W)^{D_{i}(G)}.
\end{align}
Also let $H = [g_{\ell-1}^{\rm T},\,\cdots,\,g_{0}^{\rm T}]^{-1}$. Assuming $D_i(H) \leq D_{i-1} (H)$,
\begin{align} \label{general_Z(H)}
&(1-Z(W))^{D_{i}(H)} \le 1-Z(W^{i}) \le 2^{2i+1}(1-Z(W))^{D_{i}(H)}.
\end{align}
\subsection{Proof of Theorem~\ref{main_result}}
We first provide an intuitive picture behind the result of Theorem~\ref{main_result}. For simplicity, assume $\ell=2$ and let the channel $W$ be a binary erasure channel (BEC) with erasure probability $\epsilon$. 
The capacity of this channel is $1-\epsilon$. 
For such a channel, the Bhattacharyya process has a simple closed form~\cite{Ari09} as $Z_0= \epsilon$ and 
\begin{equation} \label{Z_n}
Z_{n+1} =\left\{ \begin{array}{cc} 
Z_n^2,& B_n=0,\\
2Z_n-Z_n^2,& B_n=1.
\end{array}\right.
\end{equation}
We know from Section~\ref{related} that as $n$ grows large, 
$Z_n$ tends almost surely to a $\{0,1\}$-valued random variable $Z_{\infty}$ with $\mathbb{P}(Z_{\infty}=0)=1-\epsilon$. 
The asymptotic behavior of $\{Z_n\}$ can be explained roughly by considering the behavior of $\{-\log Z_n\}$. In particular, it is clear from \eqref{Z_n} that at time $n+1$, $-\log Z_{n}$ is either doubled (when $B_n=0$), or decreased by at most $1$ (when $B_n=1$). 
Also, observe that once $-\log Z_n$ becomes sufficiently large, subtracting $1$ from it has negligible effect compared with the doubling operation. Now assume  that $m$ is a sufficiently large number.  Conditioned on the event that $-\log Z_m$ is a very large value (or equivalently, the value of $Z_m$ is very close to $0$: this happens with probability very close to $1-\epsilon$), for $n>m$  the process $\{-\log Z_n\}$ evolves each time by being doubled if $B_n=0$ or remaining roughly the same if $B_n=1$.  
We can then use the central limit theorem to characterize the asymptotic behavior of $\{-\log Z_n\}$ for $n\gg m$.
  
The proof of  Theorem~\ref{main_result} is done by making the above intuitive steps rigorous for a BMS channel $W$ and a  polarizing $\ell \times \ell$ kernel matrix $G$. In a slightly more general setting, we study the asymptotic properties 
of $\mathbb{P}(X_n\le x)$ for any generic process $\{X_n\}_{n\in\naturals}$ satisfying the conditions (c1)--(c4) defined as follows.  
\begin{definition}\label{def:S}
Let $S$ be a random variable taking values in $[1,\infty)$.
Assume that the expectation and the variance of $\log S$ exist
and are denoted by $\mathbb{E}[\log S]$ and $\mathbb{V}[\log S]$, respectively.
Assume that $\{S_n\}_{n\in\mathbb{N}}$ are i.i.d.\ samples of $S$.
Let $\{X_n \in (0,1)\}_{n\in\mathbb{N}}$ be a random process satisfying the following conditions:
\begin{itemize}
\item[(c1)] There exists a random variable $X_\infty$ such that $X_n\to X_\infty$ holds almost surely.
\item[(c2)] $X_n^{S_n}\le X_{n+1}$.
\item[(c3)] There exists a constant $c \geq 1$ such that $X_{n+1} \le cX_n^{S_n}$ holds.
\item[(c4)] $S_n$ is independent of $X_m$ for $m\le n$.
\end{itemize}
\end{definition}
The random processes $\{Z_n\}_{n\in\mathbb{N}}$ and $\{1-Z_n\}_{n\in\mathbb{N}}$ satisfy the above four conditions
by letting $S_n = D_{B_n}(G)$ and $S_n=D_{B_n}(H)$, respectively.
The fact that these processes satisfy the condition (c1) has been proved in~\cite[Lemma 5.4]{Kor09thesis}, 
and the result reads that if $G$ is polarizing, then $Z_\infty$ takes only 0 and 1, 
with probabilities $I(W)$ and $1-I(W)$, respectively. Conditions (c2) and (c3) also hold because of \eqref{general_Z(G)} and \eqref{general_Z(H)}. 

Our objective now is to prove that for such a process $\{X_n\}_{n\in\mathbb{N}}$, we have
\begin{equation} \label{X}
\lim_{n\to\infty}
\mathbb{P}\left(X_n \le 2^{-2^{n \mathbb{E}[\log S] + t\sqrt{n\mathbb{V}[\log S]} + f(n)}}\right)
 = \mathbb{P}(X_\infty=0)Q(t),
\end{equation}
where $f(n)$ is any function such that $f(n)=o(\sqrt{n})$ holds. The results of Theorem~\ref{main_result} then follow 
by noting that $\mathbb{P}(Z_\infty=0)=I(W)$ and $\mathbb{P}(1-Z_\infty=0)=\mathbb{P}(Z_\infty=1)=1-I(W)$ hold, 
and by substituting 
$t=Q^{-1}(R/I(W))$ and $t=Q^{-1}(R'/(1-I(W)))$, respectively, into \eqref{X}. 

We prove \eqref{X} by showing 
the two inequalities obtained by replacing the equality in \eqref{X} 
by inequality in both directions.
As the first step we have:
\begin{lemma}\label{lem:main_direct}
Let $\{X_n\}_{n\in\mathbb{N}}$ be a random process satisfying (c1), (c3) and (c4).
For any $f(n)=o(\sqrt{n})$,
\begin{equation*}
\liminf_{n\to\infty}
\mathbb{P}\left(X_n \le 2^{-2^{n\mathbb{E}[\log S] + t\sqrt{n \mathbb{V}[\log S]} + f(n)}}\right)
 \ge \mathbb{P}(X_\infty=0)Q(t).
\end{equation*}
\end{lemma}
\begin{IEEEproof}
Without loss of generality, we can assume that $c$ in condition (c3) satisfies $c\ge 2$. 
Define the process $\{L_n\}_{n\in\naturals}$ as $L_n \triangleq \log X_n$.
From (c3), we have 
\begin{align*}
L_n &\le \log c + S_{n-1} L_{n-1},
\end{align*}
and by applying the above relation recursively, for $m\le n-1$ we obtain
\begin{align} \label{rec}
 L_n &\le \left(\sum_{j=m}^{n-1} \prod_{i=j+1}^{n-1} S_i\right)\log c + \left(\prod_{i=m}^{n-1}S_i\right)L_m \nonumber\\
&\le \left(\prod_{i=m}^{n-1}S_i\right)((n-m)\log c + L_m). 
\end{align}
Fix $\beta\in(0,\mathbb{E}[\log S])$ and let
\begin{equation} \label{m}
m \triangleq (\log n+\log\log c)/\beta.
\end{equation}
Conditioned on the event $\mathcal{D}_m(\beta)\triangleq \{ X_m < 2^{-2^{\beta m}}\}$, by using \eqref{rec} we obtain
\begin{equation*}
L_n\le -\left(\prod_{i=m}^{n-1} S_i\right) m\log c.
\end{equation*}
Let the event $\mathcal{H}_m^{n-1}(t)$ be defined as 
\begin{align*}
 \mathcal{H}_m^{n-1}(t)\triangleq \biggl\{ & \sum_{i=m}^{n-1}\log S_i\ge (n-m) \mathbb{E}[\log S]
+ t\sqrt{(n-m)\mathbb{V}[\log S]}+f(n-m)\biggr\},
\end{align*}
where $f$ is any function such that $f(k)=o(\sqrt{k})$ holds.
Conditioned on $\mathcal{D}_m(\beta)$ and $\mathcal{H}_m^{n-1}(t)$, we have
\begin{equation*}
\log(-L_n) \ge
\log m + \log\log c + (n-m) \mathbb{E}[\log S]\\
+t\sqrt{(n-m)\mathbb{V}[\log S]}
+f(n-m).
\end{equation*}
Hence, 
\begin{multline*}
\mathbb{P}\bigg(
\log(-L_n) \ge
\log m + \log\log c + (n-m)\mathbb{E}[\log S]
+t\sqrt{(n-m)\mathbb{V}[\log S]}
+f(n-m)\bigg)\\
\ge
\mathbb{P}(\mathcal{D}_m(\beta)\cap\mathcal{H}_m^{n-1}(t))
=\mathbb{P}(\mathcal{D}_m(\beta))\mathbb{P}(\mathcal{H}_m^{n-1}(t)).
\end{multline*}
The last equality follows from the independence condition (c4).

Note that taking the limit $n\to\infty$ also implies 
$m\to\infty$ and $n-m\to\infty$ via~\eqref{m}. 
From Theorem~\ref{thm:FirstEXP} (in Appendix), we  have 
$\lim_{n\to\infty}\mathbb{P}(\mathcal{D}_m(\beta))=\mathbb{P}(X_\infty=0)$.
We also have $\lim_{n\to\infty}\mathbb{P}(\mathcal{H}_m^{n-1}(t))=Q(t)$ 
due to the central limit theorem for $\{\log S_i\}$. 
We consequently have 
\begin{equation*}
\liminf_{n\to\infty}\mathbb{P}\bigg(
\log(-\log X_n) \ge
n \mathbb{E}[\log S] +t\sqrt{n\mathbb{V}[\log S]}
+f(n)\bigg)\\
\ge \mathbb{P}(X_\infty=0)Q(t)
\end{equation*}
for any $f(n)=o(\sqrt{n})$.
\end{IEEEproof}

The second step of the proof of \eqref{X} is to prove the other direction of the inequality. We have:
\begin{lemma} \label{lem:conv}
Let $\{X_n\}_{n\in\mathbb{N}}$ be a random process satisfying (c1), (c2) and (c4).
For any $f(n)=o(\sqrt{n})$,
\begin{equation*}
\limsup_{n\to\infty}
\mathbb{P}\left(X_n \le 2^{-2^{n\mathbb{E}[\log S] + t\sqrt{n\mathbb{V}[\log S]} + f(n)}}\right)\\
\le \mathbb{P}(X_\infty=0)Q(t).
\end{equation*}
\end{lemma}
\begin{IEEEproof}
Let $L_n\triangleq \log X_n$.
From (c2), for $m\le n-1$ we have
\begin{align*}
L_n &\ge S_{n-1} L_{n-1}\\
&\ge \left(\prod_{i=m}^{n-1}S_i\right)L_m,
\end{align*}
and thus
\begin{equation}
\log (-L_n) \le 
\sum_{i=m}^{n-1}\log S_i + \log (-L_m).\label{eq:Lupper}
\end{equation}
Hence, for any fixed $m$ and any $\delta\in(0,1)$,
\begin{align}
\nonumber&\limsup_{n\to\infty}\mathbb{P}\left(\log (-L_n) > n\mathbb{E}[\log S]+t\sqrt{n\mathbb{V}[\log S]}+f(n)\right)\\
\nonumber&\le\limsup_{n\to\infty}\mathbb{P}\bigg(\log (-L_n) > n\mathbb{E}[\log S]+t\sqrt{n\mathbb{V}[\log S]}+f(n),\; X_m\le \delta\bigg)\\
\label{eq:2terms} &\quad+\limsup_{n\to\infty}\mathbb{P}\bigg(\log (-L_n) > n\mathbb{E}[\log S]+t\sqrt{n\mathbb{V}[\log S]}+f(n),\; X_m> \delta\bigg).
\end{align}
The first term in the right-hand side of~\eqref{eq:2terms} is upper bounded as
\begin{align*}
&\limsup_{n\to\infty}\mathbb{P}\bigg(\log (-L_n)> n\mathbb{E}[\log S]+t\sqrt{n\mathbb{V}[\log S]}+f(n),\; X_m\le \delta\bigg)\\
&\stackrel{(a)}{\le}\limsup_{n\to\infty}\mathbb{P}\bigg(\sum_{i=m}^{n-1}\log S_i + \log (-L_m) > n\mathbb{E}[\log S]+t\sqrt{n\mathbb{V}[\log S]}+f(n),\; X_m\le \delta\bigg)\\
&\stackrel{(b)}{=}Q(t)\mathbb{P}(X_m\le\delta),
\end{align*}
where (a) follows from~\eqref{eq:Lupper}, and where (b) follows from (c4) and the central limit theorem.
The second term in the right-hand side of~\eqref{eq:2terms} is upper bounded as
\begin{align*}
&\limsup_{n\to\infty}\mathbb{P}\bigg(\log (-L_n)  > n\mathbb{E}[\log S]+t\sqrt{n\mathbb{V}[\log S]}+f(n),\; X_m> \delta\bigg)\\
&\le\limsup_{n\to\infty}\mathbb{P}\left(X_n\le\frac{\delta}{2},\; X_m>\delta\right)\\
&\stackrel{(a)}{\le} \mathbb{P}\left(X_\infty\le\frac{\delta}{2},\;X_m>\delta\right),
\end{align*}
where (a) follows from (c1).
Applying these bounds to~\eqref{eq:2terms}, for any $\delta\in(0,1)$, we have
\begin{align*}
&\limsup_{n\to\infty}\mathbb{P}\left(\log (-L_n) > n\mathbb{E}[\log S]+t\sqrt{n\mathbb{V}[\log S]}+f(n)\right)\\
&\le\limsup_{m\to\infty} \left\{Q(t)\mathbb{P}(X_m\le\delta) + \mathbb{P}\left(X_\infty\le\frac{\delta}{2},\;X_m>\delta\right)\right\}\\
&\le Q(t)\mathbb{P}(X_\infty\le\delta) + \mathbb{P}\left(X_\infty\le\frac{\delta}{2},\;X_\infty\ge\delta\right)\\
&=Q(t)\mathbb{P}(X_\infty\le\delta).
\end{align*}
By letting $\delta\to0$, we obtain the result.
\end{IEEEproof}

\subsection{Proof of Theorem~\protect\ref{main_result2}}
\begin{lemma}\label{lem:rep}
The MAP error probability of a linear code $\mathcal{C}$ over a BMS channel $W$ is lower bounded by
$Z(W)^{2d_\text{min}}/4$
where $d_\text{min}$ is the minimum distance of $\mathcal{C}$.
\end{lemma}
\begin{IEEEproof}
Within this proof, the notation $\mathbb{P}(\cdots)$ should be understood as 
generically denoting the probability of an event $(\cdots)$. 
Since the MAP error probability of a linear code over a BMS channel does not depend on transmitted codeword,
we can assume without loss of generality that transmitted codeword is the all-zero codeword, 
which is denoted by $\mathbf{0}$.
Let $\pmb{Y}$ be the random variable corresponding to a received sequence when $\mathbf{0}$ is transmitted
and let $P(y\mid c)$ be the likelihood of a codeword $c$ given a received sequence $y$.
Since MAP and ML are equivalent for equiprobable codewords, the MAP error probability is lower bounded as
\begin{align*}
&\mathbb{P}(\cup_{c'\in\mathcal{C}\setminus\{\mathbf{0}\}} \left\{P(\pmb{Y}\mid c') \ge P(\pmb{Y}\mid \mathbf{0})\right\})\ge
\mathbb{P}(P(\pmb{Y}\mid c) \ge P(\pmb{Y}\mid \mathbf{0}))\\
&=
P_e(W^{\otimes w(c)})\\
& \stackrel{(a)}{\ge}
\frac12 \left(1-\sqrt{1-Z(W^{\otimes w(c)})^2}\right)\\
&=\frac12 \left(1-\sqrt{1-Z(W)^{2w(c)}}\right)\\
&\ge
\frac14 Z(W)^{2w(c)}.
\end{align*}
Here, $c$ is an arbitrary codeword in the set $\mathcal{C}\setminus\{\mathbf{0}\}$
and $w(c)$ denotes its Hamming weight.
Also $W^{\otimes m}$  denotes the $m$-parallel channel of $W$ which has the following rule
\begin{equation}
W^{\otimes m}(y_1^m\mid x)\triangleq\prod_{i=1}^m W(y_i\mid x).
\end{equation}
Step (a) follows from~\eqref{eq:PeZ}.
\end{IEEEproof}
It should be noted that the lower bound $P_e(W^{\otimes w(c)})\ge (1/4) Z(W)^{2w(c)}$ in the proof of Lemma~\ref{lem:rep} is not asymptotically tight
in terms of the conventional exponents.
It is possible to obtain tighter lower bounds via more elaborate arguments as in \cite[Chapter 4] {RiU08}. However, since we are only interested in behavior of double exponents, 
the above bound turns out to be sufficient for the purpose of proving Theorem~\ref{main_result2}. 


In order to prove Theorem~\ref{main_result2}, from Lemma~\ref{lem:rep} it is sufficient to prove that given any $\epsilon>0$ there exists an integer $M \in \naturals$ such that for $n \geq M$,
\begin{equation*}
 \log_\ell (d(n,R)) \leq n E_w(G) + \sqrt{n V_w(G)}\left(Q^{-1} \left(\frac{R}{I(W)}\right)+\epsilon\right),
\end{equation*}
 where $d(n,R)$ is the minimum distance of a polar code using the kernel matrix $G$, with block-length $N=\ell^n$ and rate $R$.
Since a row weight of the generator matrix is an upper bound of the minimum distance for a linear code, 
and since the weight of the $i$th row of $G^{\otimes n}$ is equal to $\prod_{j=1}^n w_{i_j}(G)$, 
where $i_j$ is the $j$th digit of the $\ell$-ary representation of $i-1$,
it is therefore sufficient to prove that given any $\epsilon>0$, there exists an integer $M \in \naturals$ such that for a polar code of block-length $N=\ell^n \geq \ell^M$ and rate $R$ and set of chosen indices $\mathcal{I}$, there exists $i\in\mathcal{I}$ 
for which the inequality 
\begin{equation}\label{eq:ineqw}
\sum_{j=1}^n \log_\ell w_{i_j}(G) \leq n E_w(G) + \sqrt{n V_w(G)}\left(Q^{-1} \left(\frac{R}{I(W)}\right)+\epsilon\right)
\end{equation}
holds. 
In the proof of Theorem~\ref{main_result}, one can observe that the key idea is to 
apply central limit theorem for $\{\log S_n=\log D_{B_n}(G)\}_{n\in\naturals}$. 
In the same sense, in order to prove Theorem~\ref{main_result2} 
we consider the random process $\{\log w_{B_n}(G)\}_{n\in\naturals}$ 
in addition to $\{\log D_{B_n}(G)\}_{n\in\naturals}$. 
Note that these processes are in general correlated 
since they are both coupled to the same process $\{B_n\}_{n\in\naturals}$. 
These processes are equal with probability one 
in the special case where $D_i(G)=w_i(G)$ holds for all $i\in\{0,\,1,\,\ldots,\,\ell-1\}$. 
In the same manner as the proof of Theorem~\ref{main_result}, 
we move on to a more abstract setting, by introducing a random variable $U$ taking values in $[1,\infty)$, 
for which we assume that the expectation and the variance of $\log U$ exist
and are denoted by $\mathbb{E}[\log U]$ and $\mathbb{V}[\log U]$, respectively, 
and by letting $\{(S_n,U_n)\}_{n\in\mathbb{N}}$ be i.i.d.\ drawings of $(S,U)$, 
where $S$ is defined as in Definition~\ref{def:S}.
Let $\{(X_n, S_n, U_n)\}_{n\in\naturals}$ be a random process such that $\{(X_n, S_n)\}_{n\in\naturals}$ 
satisfies the conditions (c1) to (c4) together with the additional condition (c5) for $\{U_n\}_{n\in\naturals}$.
\begin{itemize}
\item[(c5)] $U_n$ is independent of $X_m$ for $m\le n$.
\end{itemize}
It is easy to see that the stochastic process of the triplets $\{(Z_n, D_{B_n}(G), w_{B_n}(G))\}_{n\in\naturals}$ 
satisfies  (c1) to (c5).
We first note from the proof of Theorem~\ref{main_result} that for any generic process $\{(X_n, S_n, U_n)\}_{n\in\naturals}$ satisfying (c1) to (c5),  the relation \eqref{X} holds for any 
function $f(n)=o(\sqrt{n})$. We also claim that for real numbers $v,t$ such that $v>t$ and for any function $g(n)=o(\sqrt{n})$ we have
\begin{multline}
\limsup_{n\to\infty}
\mathbb{P}\bigg(X_n \le 2^{-2^{n\mathbb{E}[\log S] + t\sqrt{n\mathbb{V}[\log S]} + f(n)}},\\
\sum_{i=0}^{n-1} \log U_i > n\mathbb{E}[\log U] + v\sqrt{n\mathbb{V}[\log U]} + g(n)
 \bigg)
< \mathbb{P}(X_\infty=0)Q(t).
\label{eq:map_ineq}
\end{multline}
Using the relations \eqref{X} and \eqref{eq:map_ineq} it is easy to see that for generator matrices of polar codes with rate $R$, 
the number of rows satisfying~\eqref{eq:ineqw} is asymptotically proportional to the block-length, and hence 
there exists at least a row satisfying~\eqref{eq:ineqw}. We now turn to the proof of \eqref{eq:map_ineq}. 
\begin{lemma}\label{lem:2d}
Let $\{(X_n, S_n, U_n)\}_{n\in\mathbb{N}}$ be a random process satisfying (c1) to (c5).
For any $f(n)=o(\sqrt{n})$ and $g(n)=o(\sqrt{n})$,
\begin{multline*}
\lim_{n\to\infty}
\mathbb{P}\bigg(X_n \le 2^{-2^{n\mathbb{E}[\log S] + t\sqrt{n\mathbb{V}[\log S]} + f(n)}},\,
\sum_{i=0}^{n-1} \log U_i > n\mathbb{E}[\log U] + v\sqrt{n\mathbb{V}[\log U]} + g(n)
 \bigg)\\
= \mathbb{P}(X_\infty=0)\mathbb{P}(A_S \ge t, A_U \ge v),
\end{multline*}
where $(A_S,A_U)$ are Gaussian random variables of mean zero whose covariance matrix is equal to that of
\begin{equation*}
\left(\frac{\log S - \mathbb{E}[\log S]}{\sqrt{\mathbb{V}[\log S]}},\;
\frac{\log U - \mathbb{E}[\log U]}{\sqrt{\mathbb{V}[\log U]}}
\right).
\end{equation*}
\end{lemma}
The proof of this Lemma is the same as the proofs of Lemma~\ref{lem:main_direct} and Lemma~\ref{lem:conv}.
The difference is that the central limit theorem is replaced by the two-dimensional central limit theorem.
From $\mathbb{P}(A_S\ge t, A_U\ge v)\le Q(\max\{t,v\})$, the relation \eqref{eq:map_ineq} is obtained for $v>t$. This completes the proof of Theorem~\ref{main_result2}.

\emph{Remark}:  Let $G=\bigl[ \begin{smallmatrix} 1 &0 \\ 1& 1 \end{smallmatrix}\bigl]$.  For this choice of $G$, we have $w_i(G) = D_i(G)$ for $i = 0$ and $1$. Hence, the random variables $S_n= D_{B_n}(G)$ and $U_n=w_{B_n}(G)$ are equal for $n \in \naturals$. Also note that $S_n$ takes its value in the set $\{1,2\}$ uniformly at random. From the proof of Theorem~\ref{main_result2}, 
the set of indices of the rows of polar codes with the kernel matrix $G$ and rate $R$ correspond to the event
\begin{equation*}
\left\{X_n \le 2^{-2^{n\mathbb{E}[\log S] + Q^{-1}\left(\frac{R}{I(W)}\right)\sqrt{n\mathbb{V}[\log S]} + f(n)}}\right\}.
\end{equation*}
Also, with the same $G$,  the set of indices of a RM code with rate $R'$ correspond to the event
\begin{equation*}
\left\{\sum_{i=0}^{n-1} \log U_i > n\mathbb{E}[\log U] + Q^{-1}(R')\sqrt{n\mathbb{V}[\log U]} + g(n)\right\}.
\end{equation*}
From Lemma~\ref{lem:2d}, it is easy to conclude that
the fraction of the common chosen row indices of $G^{\otimes n}$ between polar codes of 
rate $R$ and RM codes of rate $R'$ tends to  $I(W)\min\{\frac{R}{I(W)},R'\}$ as $n \to \infty$.

\section{Selection rule of rows} \label{select_sec}
The proof of Lemma~\ref{lem:main_direct}
suggests a way to help us select the good indices in a more computationally efficient way. 
In the proof, $\ell$-ary expansion of row indices of $G^{\otimes n}$ corresponds to realizations of $B_1,\dotsc,B_n$.
The proof of Lemma~\ref{lem:main_direct} implies that 
it is sufficient to select rows in $\mathcal{D}_m(\beta)\cap\mathcal{H}_m^{n-1}(t)$ 
in order to achieve the asymptotically optimum performance.
It should be noted that the event $\mathcal{D}_m(\beta)$ applied to the Bhattacharyya process 
$\{Z_n=Z(W_n)\}_{n\in\naturals}$ of $W$ depends on the channel $W$, whereas the event $\mathcal{H}_m^{n-1}(t)$ is channel-independent. 
This observation leads to the following selection rule:
The first $m=s(n)\triangleq(\log n + \log\log c)/\beta$ digits of the row indices are determined in the channel-dependent way.
Then, the following $(n-m)$ digits are determined in the RM way, i.e., 
those combinations of digits $(B_m,\,\dotsc,\,B_{n-1})$ giving large values of $\sum_{i=m}^{n-1}\log D_{B_i}(G)$ are selected.
In this rule, only the first $\Theta(\log n)$ digits should be determined depending on the channel.

The above argument can further be extended in a recursive manner. 
Let $\mathcal{C}_m^{n-1}(\epsilon) \triangleq \{(n-m)^{-1}\sum_{i=m}^{n-1}\log S_i\ge \mathbb{E}[\log S] -\epsilon\}$. 
Then, it is sufficient to select rows in
$\mathcal{D}_{m_0}(\beta)\cap\mathcal{C}_{m_0}^{m_1-1}(\epsilon)\cap\mathcal{H}_{m_1}^{n-1}(t)$
where $m_1=s(n)$ and $m_0=s(m_1)$ since $\mathcal{D}_{m_1}(\beta)$ and $\mathcal{D}_{m_0}(\beta)\cap \mathcal{C}_{m_0}^{m_1-1}(\mathbb{E}[\log S] - \beta)$ are asymptotically equal.
(Use $\mathcal{C}_m^{n-1}(\epsilon)$ instead of $\mathcal{H}_m^{n-1}(t)$ in the proof of Lemma~\ref{lem:main_direct}. A similar argument can be found in~\cite[Section IV-B]{Ari09}.) 
From this observation, only $\Theta(\log\log n)$ digits have to be determined depending on the channel.
By iterating this argument, we obtain the selection rule in which only 
\begin{equation} \label{select}
\Theta(\overbrace{\log\dotsm\log}^{k} n)
\end{equation}
digits depend on the channel for any $k\in\mathbb{N}$.
From the argument so far, we deduce that even though the behavior of $Z_n=Z(W_n)$ depends 
on the channel $W$ as well as the whole sequence $\{B_0,\,B_1,\,\ldots,\,B_{n-1}\}$, 
the ``fate'' regarding whether it approaches 0 or 1 when $n$ is large, is mostly determined 
by the channel $W$ and a prefix of $\{B_0,\,B_1,\,\ldots,\,B_{n-1}\}$ with a relatively small length. Thus,  to choose the  indices of the 
 channels $W_{\ell^n}^{(i)}$ that have the best quality,  the first
sublinear number of significant bits of the $\ell$-ary expansion of $i - 1$ are determined 
depending on the channel and the rest are determined in a RM-like fashion.
It should be noted that the above argument is valid in the large-$n$ asymptotics. 
It does not mean that one can make the number of digits to be determined 
in the channel-dependent manner arbitrarily small. 

Although the good indices of the rows of $G^{\otimes n}$  can be selected using density evolution~\cite{MT09},
in practice storage and convolution of probability density functions is exponentially (in block-length $N$)
 costly in terms of memory and computation.
Recently, several authors have considered accurate and efficient implementation of 
the density evolution procedure~\cite{TV}, \cite{RHTT}.
The above-mentioned construction rule can be useful in reducing the number of convolutions and the number of levels in the quantization of channels.

\appendix

\begin{theorem}\label{thm:FirstEXP}
Let $\{X_n\in(0,\,1)\}_{n\in\mathbb{N}}$ be a random process satisfying (c1) and (c3).
For any fixed $\beta\in(0,\mathbb{E}[\log S])$,
\begin{equation*}
\lim_{n\to\infty} \mathbb{P}\left(X_n \le 2^{-2^{\beta n}}\right) = \mathbb{P}(X_\infty=0).
\end{equation*}
\end{theorem}
{\em Remark:} Although Theorem~\ref{thm:FirstEXP} has already been stated for 
Bhattacharyya processes $\{Z_n\}_{n\in\naturals}$ in~\cite{AT09,Kor09thesis},
we would nevertheless like to confirm that the result is obtained by using
only the two conditions (c1) and (c3).
\begin{IEEEproof}[Proof of Theorem~\ref{thm:FirstEXP}]
As the inequality 
\begin{equation*}
\limsup_{n\to\infty} \mathbb{P}\left(X_n \le 2^{-2^{\beta n}}\right) \le \mathbb{P}(X_\infty=0)
\end{equation*}
obviously holds, a proof of the  lower bound
\begin{equation*}
\liminf_{n\to\infty} \mathbb{P}\left(X_n \le 2^{-2^{\beta n}}\right) \ge \mathbb{P}(X_\infty=0)
\end{equation*}
is given in the following.
Fix $\epsilon\in(0,1)$.
Let $\{J_n\}_{n\in\mathbb{N}}$ be the random process defined as
\begin{equation*}
J_n \triangleq 
\left\{
\begin{array}{ll}
\log(-\log X_n),&\hspace{2em} \text{for } n = 0,\dotsc,m\\
\log(S_{n-1} - \epsilon) + J_{n-1},&\hspace{2em} \text{for } n>m,
\end{array}\right.
\end{equation*}
which is to be used for deriving a probabilistic bound for $\{X_n\}_{n\in\naturals}$. 
Let $\mathcal{T}_m^n(\gamma)\triangleq\{X_i < \gamma,\text{ for } i = m,m+1,\dotsc,n\}$.
Fix $k\in\{1,\,2,\,\ldots\}$.
From (c3), conditioned on $\mathcal{T}_m^{m+k-1}(c^{-1/\epsilon})$, the inequality $\log(-\log X_n) \ge J_n$ holds
for $n=m,m+1,\dotsc,m+k$.
For the process $\{J_n\}_{n\in\naturals}$, the inequality 
\begin{align*}
J_{m+k}&=J_m+\sum_{i=m}^{m+k-1}\log(S_i-\epsilon)\\
&\ge J_m + \sum_{i=m}^{m+k-1}(\log S_i+\log(1-\epsilon))
\end{align*}
holds since $S_i\ge 1$. 
This inequality immediately implies the following conditional bound:
Conditioned on $\mathcal{C}_m^{m+k-1}(\epsilon) \triangleq \{(1/k)\sum_{i=m}^{m+k-1}\log S_i\ge \mathbb{E}[\log S] -\epsilon\}$, one has 
\begin{equation*}
J_{m+k}\ge J_m + k(\mathbb{E}[\log S] - \epsilon + \log(1-\epsilon)).
\end{equation*}
We have therefore obtained a probabilistic bound of $\log(-\log X_{m+k})$ of the form 
\begin{multline*}
\mathbb{P}(\log(-\log X_{m+k}) \ge J_m+k(\mathbb{E}[\log S] - \epsilon + \log(1-\epsilon)))\\
\ge \mathbb{P}\left(\mathcal{T}_m^{m+k-1}(c^{-1/\epsilon})\cap\mathcal{C}_m^{m+k-1}(\epsilon)\right)\\
\ge \mathbb{P}\left(\mathcal{T}_m^{m+k-1}(c^{-1/\epsilon})\right) + \mathbb{P}\left({\mathcal{C}_m^{m+k-1}}(\epsilon)\right) -1,
\end{multline*}
for any $m\in\mathbb{N}$, $k\in\mathbb{N}$ and $\epsilon>0$.
From the law of large numbers, $\lim_{k\to\infty} \mathbb{P}\left({\mathcal{C}_m^{m+k-1}}(\epsilon)\right) = 1$.
From (c1), $\lim_{m\to\infty}\lim_{k\to\infty} \mathbb{P}\left(\mathcal{T}_m^{m+k-1}(c^{-1/\epsilon})\right) \ge \mathbb{P}(X_\infty < c^{-1/\epsilon})$.
Hence,
\begin{align*}
&\liminf_{m\to\infty}\liminf_{k\to\infty}\\
&\mathbb{P}(\log(-\log X_{m+k}) \ge J_m+k(\mathbb{E}[\log S] - \epsilon + \log(1-\epsilon)))\\
&\hspace{8em}\ge \mathbb{P}(X_\infty < c^{-1/\epsilon})\ge\mathbb{P}(X_\infty = 0)
\end{align*}
holds for any $\epsilon>0$.
On the other hand, we observe that 
\begin{align*}
&\liminf_{n\to\infty} \mathbb{P}\left(\frac1n\log(-\log X_n)\ge\mathbb{E}[\log S] - \gamma\right)\\
&\ge\liminf_{k\to\infty}\\
&\mathbb{P}(\log(-\log X_{m+k}) \ge J_m+k(\mathbb{E}[\log S] - \epsilon + \log(1-\epsilon)))
\end{align*}
holds for any fixed $m\in\mathbb{N}$ and $\gamma > \phi(\epsilon) \triangleq \epsilon-\log(1-\epsilon)$.
Hence,
\begin{align*}
&\liminf_{n\to\infty} \mathbb{P}\left(\frac1n\log(-\log X_n)\ge\mathbb{E}[\log S] - \gamma\right)\\
&\quad\ge \mathbb{P}(X_\infty = 0)
\end{align*}
for any $\gamma >0$ since $\phi(\epsilon) > 0$ for $\epsilon>0$ and $\lim_{\epsilon\to0}\phi(\epsilon)=0$.
\end{IEEEproof}

\end{document}